# The joint optimization of critical interdependent infrastructure of an electricity-water-gas system

Jie Cheng[ab], Qishuai Liu[a], Qing Hui[a] *, Fred Choobineh[a]

[a]*University of Nebraska-Lincoln, SEC 209N, Lincoln 68588, USA*
[b]*CalEnergy Operating Corporation, 7030 Gentry Rd, Calipatria 92233, USA*

**Abstract**

Electricity, water, and gas systems are critical infrastructures that are sustaining our daily lives. This paper studies the joint operation of these systems through a proposed optimization model and explores the advantage of considering the system of systems. Individual and joint optimizations are studied and compared. The numerical results show that the total electricity cost for these three systems can be reduced by 9% via joint optimization. Because the water system and gas system intrinsically include the storages in their systems, the power system can use these storages as the regulation capacity to shift load from peak hours to off-peak hours. Since the saving on the power generation cost surpasses the incremental cost in the operation and maintenance (O&M), the overall economic performance is improved by the joint optimization.
© 2018 The Authors.

**Nomenclature**

<u>Index and constant</u>

| | |
|---|---|
| $a_n, b_n$ | Parameters for pricewise linearization in kW and $, respectively. |
| $c_1, c_2, c_3$ | Cost coefficients for quadratic electric power curve in $/kW$^2$, $/kW, and $, respectively. |
| $h_g$ | Power coefficient for gas load flow in kW/(m$^3$/h). |
| $h_w$ | Power coefficient for water load flow in kW/(m$^3$/h). |
| $N^s$ | Number of points for piecewise linearization. |
| $N^T$ | Number of time steps. |
| $p_p^{ref}$ | Nominal pipe pressure reference in Pa. |
| $r_e'$ | Pseudo electricity rate in $/kWh. |
| $r_e$ | Finalized electricity rate in $/kWh. |

* Corresponding author. Tel: (402) 472-3714; fax: (402) 472-4732.
  *E-mail address:* qing.hui@unl.edu

| | |
|---|---|
| $r_g$ | O&M cost coefficient for gas storage system in $/m$^3$. |
| $r_p$ | O&M cost coefficient for pipe system in $/Pa. |
| $r_s$ | Coefficient of Per-unit cost of gas supply in $/unit. |
| $r_w$ | O&M cost coefficient for water system in $/m$^3$. |
| $S_p^{ref}$ | Nominal pipe storage reference in m$^3$. |
| $t$ | Time index. |
| $T$ | Time step constant in hour. |
| $V_g$ | Volume of one unit gas in gas transportation in m$^3$/unit. |

Variables

| | |
|---|---|
| $\lambda_n(t)$ | Ancillary variable for pricewise linearization. |
| $L_r(t)$ | Residential electricity load at time *t* in kW. |
| $L_g(t)$ | Gas load at time *t* in m$^3$/h. |
| $L_w(t)$ | Water load at time *t* in m$^3$/h. |
| $m(t)$ | Gas transportation decision variable at time *t* in per unit. |
| $p_p(t)$ | Pipe Pressure status at time *t* in Pa. |
| $P_e(t)$ | Residential electric load (non-infrastructure electric load) at time *t* in kW. |
| $P_g(t)$ | Gas system electric load at time *t* in kW. |
| $P_w(t)$ | Water system electric load at time *t* in kW. |
| $Q_g(t)$ | Gas flow rate at time *t* in m$^3$/h. |
| $Q_w(t)$ | Water flow rate at time *t* in m$^3$/h. |
| $S_g(t)$ | Gas tank storage status at time *t* in m$^3$. |
| $S_p(t)$ | Gas pipe storage status at time *t* in m$^3$. |
| $S_w(t)$ | Water storage status at time *t* in m$^3$. |

## 1. Introduction

The national critical infrastructures are those that provide the essential services and serve as the backbone of the nation's economy, security, and health. The proper functioning and coordination of these infrastructures is essential for the nation's economic development and security since these infrastructure caters to the basic needs of the population.

The electricity, water, and gas (EWG) systems are among the most critical infrastructures of a community, and traditionally each has been planned, designed and operated isolated from each other. Most studies and optimization of these systems have been done individually and within the scope of a homogeneous system.

The homogeneous system optimization can be found in the following literatures. The optimal power flow is optimized by the distributed algorithms based on the multi-agent system[1]. The models and methods of optimal control of water distribution systems is summarized in[2]. The optimal operation of water distribution is studied in[3], considering the water shortage and the quality constraints. A stochastic optimal control frame for natural gas network operation is proposed in[4]. Genetic algorithm and model predictive control are employed to solve the optimal operation of pipelines in[5] and[6], respectively.

However, the EWG systems are functionally and economically linked. For example, water and gas system use electrical energy for its operation, and electrical system may rely on water for its cooling and gas as a source of fuel. Moreover, the EWG systems not only share some common features but they also share some complementary characteristics. The power system is a real-time balanced system, where at any given time the generation is equated with the load and the system does not have the benefit of a storage in the transient form. The water system is an asynchronous system, where the water storage that provides water pressure by height naturally provides a storage buffer for the gap between the water production and the water load. The natural gas is compressible in the pipelines.

It forms a big inertia system to absorb supply fluctuation within a certain range. Thus, it is insufficient to only consider one system within the scope of optimization, since these infrastructures are interdependent.

Some bilateral interdependent system study can be found in[7,8,9]. However, tri-lateral interdependent systems are rarely studied.

This paper emphasizes on the energy linkage and load balance among/between the three systems. The power, water, gas system will be modeled and optimized separately first as a benchmark. Then, the joint economic operation of an EWG system as a whole system is studied. Finally, the economic performance is compared.

## 2. Critical Infrastructure

The homeland security defines the critical infrastructure as the infrastructure that provides the essential services underpinning the American society and serves as the backbone of our nation's economy, security, and health[10]. The electricity, gas, oil, water, transportation, and communication systems are the common critical infrastructure that are associated with and have fundamentally shaped our daily lives. Other less tangible critical infrastructure includes the public health, food safety, financial services, and security services (police, military).

We selected the EWG system because they share some common features, such as the natural monopoly in one area, non-sufficient competition, major energy consumers, critical element to the residents, continuous supply in physical networks, simultaneous load balance, and no storage in the end users. The power, water, and gas distribution systems are introduced as follows, respectively.

### 2.1. Power System

The power system includes the power generation, transmission and distribution, and the load. The power is generated by various generation resources, and transmitted through the transmission and distribution networks, and reaches to the end users. The electricity is a real-time commodity, where the generation and consumption happen simultaneously.

In this paper, the aggregated model is used to represent generation and load. The generation from different sources and the load from different end users are aggregated as one bulk generator and on bulk load. Thus, a single generator and a single load can represent the behavior of the power supply and power consumption pattern in the represented area.

### 2.2. Water System

The water system, includes the water sources/reservoir, treatment plant, pump station, storage tank/tower, distribution network, and commercial and residential loads. The water level in the water tower has to be maintained in a certain range in order for the outlet water pressure be maintained in the required pressure range for the customers. The single provider and one aggregated load model are used in this system.

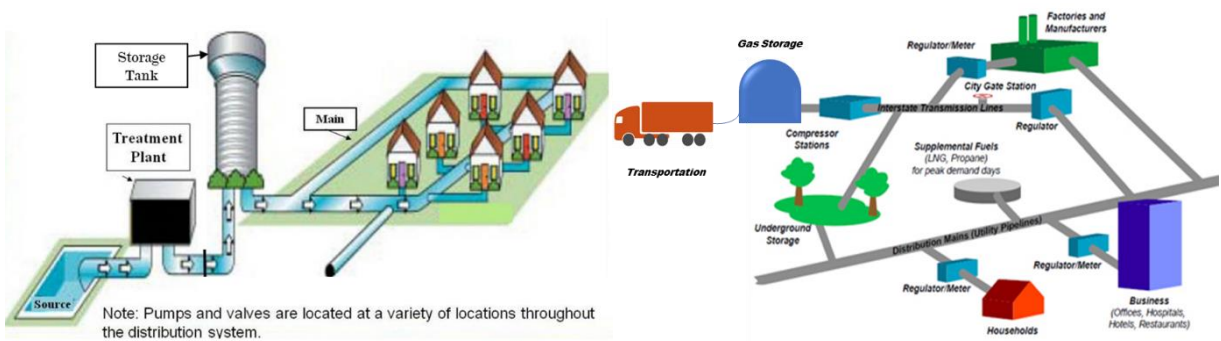

Fig 1. Layout of (a) water system network and (b) gas system network[11,12].

*2.3. Gas System*

The gas system includes the gas transportation, tank storage, compression station, distribution pipelines, and commercial and residential end users.

Because the gas is compressible, the pipelines contain a certain amount of natural gas that can be considered as another type of storage. Gas pressure has to be maintained within the allowable range for the satisfaction of end users. The single provider and one aggregated load model are used in this system.

**3. Separate and Joint Optimizations**

In order to analyze the performance of different operation strategies of these systems, two cases are considered for simulation.

Case 1) Independent system operation.

This case assumes that the three systems are operated independently. The system models are constructed and optimized, separately. The objective of the optimization is to minimize the respective system cost consisting of electricity-related energy cost and operation and maintenance (O&M) costs.

A pseudo electricity rate is introduced to facilitate the optimization of the O&M cost of the water and gas distribution system. Then, water and gas system electric load (infrastructure electric load) and the residential electric load (non-infrastructure electric load) are aggregated in each hour and form a load curve. The total electric cost (infrastructure + non-infrastructure) is divided by the total amount of the electricity to obtain the finalized flat power rate. Finally, the finalized flat electricity rate is used to substitute the pseudo electricity rate to re-calculate the real cost of the water and gas systems.

Case 2) Interdependent optimization.

This case inherits the same balance functions and capacity constraints constructed in Case 1. The difference is the objective function will not be minimized until all the constraints are formulated and listed.

Then, the problem is optimized with the programming method, and the final results give the breakdowns of the optimal cost values and the corresponding operation strategies. The finalized electricity rate will be calculated as a byproduct.

*3.1. Water balance optimization*

The water storage balance is described in (1), where the storage status for the next time step is the existing water balance, plus flow-in, minus flow-out in the same period. The electric power for the corresponding flow rate is depicted in (2). In order to simulate the sustainable cycle of water balance, the final state of the storage should be equal to the initial state of the storage. The storage capacity and the flow capacity should meet their respective limits. In the water system optimization, the known variable is the water load for $N^T$ time steps, the decision variable is the water flow into the storage at time *t*. The objective function consists of the electric cost and the O&M cost as shown in (4). The problem can be solved by the linear programming method.

$$S_w(t) = S_w(t-1) + Q_w(t) \cdot T - L_w(t) \cdot T \tag{1}$$

$$P_w(t) = h_w \cdot Q_w(t) \tag{2}$$

$$t \in \{0,1,2,3, \ldots \ldots, N^T\} \tag{3}$$

$$\text{Min } Z_w = \sum_{i=1}^{N^T} r'_e \cdot h_w \cdot Q_w(t) + \sum_{i=1}^{N^T} r_w \cdot S_w(t) \tag{4}$$

*3.2. Gas balance optimization*

The gas storage balance and pipe storage balance are described in (5)-(6), respectively. The electric power for the corresponding gas flow rate is depicted in (7). The relationship between the pipe gas pressure and pipe storage status is depicted in (8). It is noted that the O&M cost for the pipe is proportional to the gas pressure in the pipe, which is an indirect relationship to the status of pipe storage.

In the gas system optimization, the known variable is the gas load vector. The decision variables are the transportation decision variable and the gas flow vector (time series). It is noted that only integer number of gas transportation is allowed in this study.

The objective function consists of the power cost, the gas transportation cost, and the O&M costs for both storage and the pipe, as shown in (10). The problem can be solved by the mixed integer linear programming method.

$$S_g(t) = S_g(t-1) + m(t) \cdot V_g \cdot T - Q_g(t) \cdot T \tag{5}$$

$$S_p(t) = S_p(t-1) + Q_g(t) \cdot T - L_g(t) \cdot T \tag{6}$$

$$P_g(t) = h_g \cdot Q_g(t) \tag{7}$$

$$p_p(t) = S_p(t) \frac{p_p^{ref}}{S_p^{ref}} \tag{8}$$

$$m(t) \in \text{non-negtive integer} \tag{9}$$

$$\text{Min } Z_g = \sum_{i=1}^{N^T} r_e' \cdot h_g \cdot Q_g(t) + \sum_{i=1}^{N^T} r_s \cdot m(t) + \sum_{i=1}^{N^T} r_g \cdot S_g(t) + \sum_{i=1}^{N^T} r_p \cdot p_p(t) \tag{10}$$

*3.3. Power balance*

The aggregated electric demand is depicted in (11), consisting of water electric load, gas electric load and the residential electric load (non-infrastructure electric load). The generation capacity constraint is described in (12). The single representative generation model is used, and the coefficients of the quadratic cost function are given as the known parameters $c_1$, $c_2$, and $c_3$. Total electric load is the summation of infrastructure power load and non-infrastructure power load. The total electric cost is described in (13). The finalized electricity rate is calculated in (14).

$$P_e(t) = h_w \cdot Q_w(t) + h_g \cdot Q_g(t) + L_r(t) \tag{11}$$

$$0 \leq P_e(t) \leq P_e^{UP} \tag{12}$$

$$\text{Min } Z_e = \sum_{i=1}^{N^T} c_1 \cdot P_e^2(t) + c_2 \cdot P_e(t) + c_3 \tag{13}$$

$$r_e = \frac{Z_e}{\sum_{i=1}^{N^T} P_e(t)} \tag{14}$$

*3.4. Joint optimization*

The joint optimization inherits all previous balance equations and the constraints but not the objective functions. The pseudo electricity rate will not be used because the finalized electricity rate will be calculated as a byproduct of the programming. The water and gas systems will not execute the optimization separately because the total electric cost will be optimized in the aggregated objective function as shown in (15).

$$\text{Min } Z_T = \sum_{i=1}^{N^T} c_1 \cdot P_e^2(t) + c_2 \cdot P_e(t) + c_3 + \sum_{i=1}^{N^T} r_w \cdot S_w(t) + \sum_{i=1}^{N^T} r_s \cdot m(t) + \sum_{i=1}^{N^T} r_g \cdot S_g(t) + \sum_{i=1}^{N^T} r_p \cdot p_p(t) \tag{15}$$

The joint optimization involves the nonlinear term in its objective function, which may introduce difficulty in problem-solving. A piecewise linearization technique is used to linearize the objective function.

Assume vector $a_n$ is the index setting points in the quadratic cost curve, and vector $b_n$ is the value points on the curve. By introducing the ancillary variable $\lambda_n(t)$, the objective function can be expressed by (20). Additional constraints[13] are described in (16)-(19). Then the formulation only includes the linear constraints and objective function. The problem can be solved by the mixed integer linear programming method.

$$0 \leq a_n \leq P_e^{UP} \tag{16}$$

$$b_n = c_1 \cdot a_n^2 + c_2 \cdot a_n + c_3 \tag{17}$$

$$n \in \{0,1,2,3,\ldots\ldots,N^s\} \tag{18}$$

$$\sum_{n=1}^{N^s} \lambda_n(t) = 1 \; for \; all \; t \tag{19}$$

$$\text{Min } Z_e = \sum_{i=1}^{N^T} \sum_{n=1}^{N^s} b_n \cdot \lambda_n(t) + \sum_{i=1}^{N^T} r_w \cdot S_w(t) + \sum_{i=1}^{N^T} r_s \cdot m(t) + \sum_{i=1}^{N^T} r_g \cdot S_g(t) + \sum_{i=1}^{N^T} r_p \cdot p_p(t) \tag{20}$$

## 4. Numerical Simulation Study

*4.1. Conditions and Parameters*

The water load, gas load, and residential electric load are obtained from the prior research works in the area of EWG systems [14,15,16] and normalized respectively by scaling down to the range to match a city load like the size of Lincoln, Nebraska. The curve is shown in Fig.2.

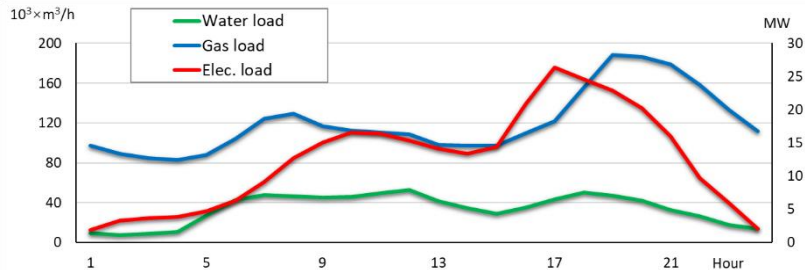

Fig 2. Water load, gas load and residential electricity load.

The simulation horizon is 24 hours, where 1-hour time steps are considered. Pseudo electricity rate is assumed to be 0.25 $/kWh. The per unit volume of gas transportation is 500 m$^3$ if converted into the normal supply status in the pipeline.

*4.2. Result and Analysis*

The power generation curves in two cases are compared in Fig.3(a), where the joint optimization significantly reduced the power fluctuation by shifting portion of the electric load from peak hours to off-peak hours.

The water power load and gas power load in two cases are compared in Fig.3(b) and (c). It is observed that since hour 16-20 are peak hours, the water and gas load shifted a portion of the loads from peak hours to off-peak hours in joint optimization. This means the joint optimization allows the water and gas systems to perceive the power cost signal, and adjust its operation strategy accordingly.

The states of the water storage, the gas tank storage, and the gas pipe storage are compared in Fig.3(d), (e) and (f). It can be seen that the average storage in the water tank and gas pipe are increased, compared to the independent optimization. This illustrates that the power system utilizes the storage facilities in the water and gas systems to redistribute its load through the time axis.

Because the water storage state is proportional to the water pressure, the simulation result also indicates the residence will have improved water pressure as a byproduct of the joint optimization. The situation is similar to the gas system, where an improved gas pressure is achieved by the co-optimization.

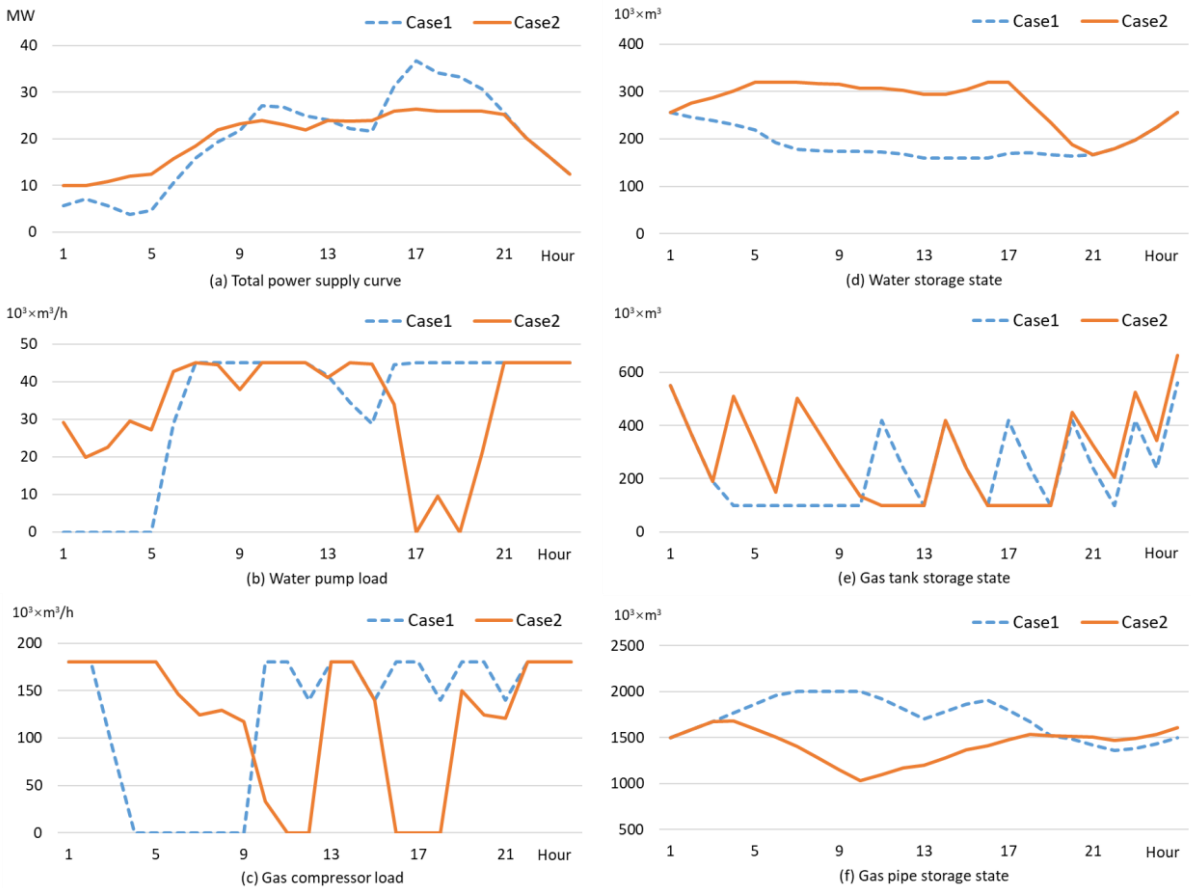

Fig 3. Comparison of a) total power supply curve, b) water electric load, c) gas electric load, d) water storage, e) gas tank storage, and f) gas pipe storage.

The explicit comparison of the components of entire system cost is given in Table 1. It is found that the O&M costs for the water and gas systems are increased by 47.2% and 12.4%, respectively. However, the electric costs are reduced by 12.7% and 15.7%, respectively. And, the residential electric cost is reduced by 12.7%.

The total EWG cost is calculated as the summation of all the O&M cost and electric cost above, i.e., A1+A2+B1+B2+B3 in Table 1. These factors lead to a reduction on the total EWG cost by 11.3%, and the finalized electricity rate is therefore reduced from 0.255$/kWh to 0.222$/kWh.

Table 1. Total cost breakdown and comparison

|  | Case1 | Case2 | Rate-of-change |
| --- | --- | --- | --- |
| Water O&M cost (A1) | $810 | $1,193 | 47.2% |
| Gas O&M cost (A2) | $7,107 | $7,986 | 12.4% |
| Water electric cost (B1) | $30,672 | $26,792 | -12.7% |
| Gas electric cost (B2) | $16,024 | $13,507 | -15.7% |
| Residual load cost (B3) | $76,434 | $66,764 | -12.7% |
| Total EWG cost | $131,048 | $116,243 | -11.3% |
| Final electric rate | $0.255 | $0.222 | -12.8% |

## 5. Conclusion

The numerical study showed the reduction on both the finalized electricity rate and the total EWG system cost. This is because the water and gas systems intrinsically included the storages in their networks. The joint optimization utilizes these resources to shift and redistribute the load through the time axis. The utilization of spare capacity of storage helped to reduce the power generation cost at the expense of a slight increase in O&M cost.